\documentclass[12pt,fleqn]{article}
\input{epsf}
\usepackage{amsmath}
\usepackage{latexsym}
\usepackage{amssymb}

\newlength{\dinwidth}
\newlength{\dinmargin}
\newcommand{\f}[2]{\frac{#1}{#2}}

\newcommand{\nn}{\nonumber}
\def\as{\alpha_S}
\def\d{\partial}
\def\hsp{\hspace{0.5mm}}
\def\bks{\hspace{-0.5mm}}
\def\bksm{\hspace{-2mm}}

\def\eps{\epsilon}
\def\z{{\tilde z}}

\def\simle{\ \lower -2.5pt\hbox{$<$} \hskip-8pt \lower 2.5pt \hbox{$\sim$}\ }
\def\simge{\ \lower -2.5pt\hbox{$>$} \hskip-8pt \lower 2.5pt \hbox{$\sim$}\ }

\begin{document}
\thispagestyle{empty}
\vspace*{-3cm}
\begin{flushright}
UPRF-98-002\\DFF 299/2/98\\
June 1998
\end{flushright}
\vspace*{2.5 cm}

\begin{center}
{\huge Fracture Functions and Jet Calculus}
\vskip 1.0cm
\begin{large}
{Gianni~Camici}\\
\end{large}
\vskip .2cm
{\it Dipartimento di Fisica,Universit\`a
 di Firenze,\\ INFN Sezione di Firenze, 50125 Firenze, Italy\\}
\vskip .7cm
\begin{large}
 {Massimiliano~Grazzini and Luca~Trentadue}\\
\end{large}
\vskip .2cm
 {\it Dipartimento di Fisica, Universit\`a di Parma,\\
 INFN Gruppo Collegato di Parma, 43100 Parma, Italy}\\
\end{center}
\vspace*{.5cm}
 \begin{abstract}
 By using Jet Calculus
 as a consistent framework to describe
 multiparton dynamics
 we explain
 the peculiar evolution equation of fracture functions
 by means 
 of the recently introduced extended fracture functions.
 \end{abstract}
\vspace*{1cm}
\begin{center}
PACS 13.85.Ni
\end{center}

\newpage
\setcounter{page}{1}

\section{Introduction}

Fracture functions \cite{tv} have been introduced
to interpret within the framework of perturbative QCD semi-inclusive
deep inelastic
processes
in the target fragmentation region.
The formulation of
these processes
by means
of fracture functions
does allow to extend the interpretation in terms of
QCD-improved parton model to the complementary region of target fragmentation.
This implies a description of deep inelastic processes in terms of the novel fracture functions
in addition to the usual factorizable
structure and fragmentation functions convoluted with hard point-like cross sections.

The introduction of these new objects requires however
the formulation of an additional factorization hypothesis whose validity
has been recently argued
on the basis of
the cut vertex formalism together with the use of infrared power counting.
These results have been obtained
within the theoretical framework of
$(\phi^3)_6$ field theory \cite{gtv,paper} and have been later confirmed in QCD in Ref.\cite{collins}.

In order to show that the new factorization hypothesis
is correct,
in Ref.\cite{gtv}
new objects have been defined called extended fracture functions,
which depend also on the momentum transfer $t$ between the incoming and outgoing hadron.
By an explicit one-loop calculation it has been verified that these object
do factorize and it has been also observed
that they show a new logarithmic dependence on the ratio of the scales
$Q^2$ and $t$ \cite{graz}. As a result extended fracture functions obey an evolution equation different from the one followed by ordinary
fracture functions.

The aim of this note is to show
in a simple and direct way
that extended and ordinary fracture functions
are closely connected. As a consequence the corresponding evolution equation
do follow one from the other.
We work within the framework of
Jet Calculus \cite{jet} by applying the corresponding rules to the evolution equations.
The set of Jet Calculus rules allows the interpretation in terms of
QCD-improved parton model of the results
obtained by using a different approach \cite{gtv}.
By using this method we can give in fact a definition of the extended
fracture function in the region where  $t$ is a perturbative scale
($\f{\as(t)}{2\pi}\simle 1$).
As a consequence a DGLAP evolution equation for extended fracture functions follows and
the inhomogeneous evolution equation
for the ordinary fracture function is recovered.
       
\section{Evolution pattern}
\label{EP}

In a deep inelastic semi-inclusive reaction, a hadron $A$ with momentum $p$
is struck by a far off-shell spacelike photon with momentum $q$
and a hadron $A^\prime$ with 
momentum $p^\prime$ is inclusively observed in the final state.
Let us define as usual
\begin{equation}
Q^2=-q^2~~~x=\f{Q^2}{2pq}
\end{equation}
and choose a frame in which $p=(p_+,p_-,{\bf 0})$ with $p_+\gg p_-$
and $pq\simeq p_+q_-$.

As far as we keep away from the target fragmentation region
the cross section factorizes as follows \cite{aemp}
\begin{equation}
\label{current}
\sigma_J=\int \f{dx^\prime}{x^\prime} \f{dz^\prime}{z^\prime}
\hsp F_A^i(x^\prime,Q^2)\hsp {\hat \sigma}_{ij} (x/x^\prime,\z/\z^\prime,Q^2)
\hsp D^j_{A^\prime}(\z^\prime,Q^2)
\end{equation}
where $F_A^i(x,Q^2)$ and $D_{A^\prime}^j (\z,Q^2)$ are the
structure
and fragmentation function respectively,
${\hat \sigma}_{ij}(x,\z,Q^2)$ is the hard semi-inclusive cross section and
we have defined $\z=pp^\prime/pq\simeq p^\prime_-/q_-$ \cite{aemp}.

Eq. (\ref{current}) expresses the fact that, as long as the produced hadron
has a large transverse momentum $p_T^{\prime 2} \sim Q^2$ (i.e. $\z$ is finite)
it can be thought as a product of the fragmentation
of the active parton.
However in the last few years,
especially after the appearance of diffractive deep
inelastic events at HERA, particular attention
has been payed to hadron production in the
target fragmentation region (i.e. the region $\z\to 0$), where eq. (\ref{current}) fails.
In Ref.\cite{tv} it has been proposed for target fragmentation 
the additional factorized term
\begin{equation}
\sigma_T=\int \f{dx^\prime}{x^\prime}
\hsp M^i_{AA^\prime}(x^\prime,z,Q^2)
\hsp {\hat \sigma}_i(x/x^\prime,Q^2)
\end{equation}
where $z=p^\prime q/pq\simeq p^\prime_+/p_+$
represent the momentum fraction of the hadron $A^\prime$ with respect to $A$.
Here $M^i_{AA^\prime}(x,z,Q^2)$ is the
fracture function, giving the probability of finding a parton
$i$ with momentum fraction $x$ in the hadron $A$ while another hadron
$A^\prime$ with momentum fraction $z$ is detected.

The fracture function
is expected to satisfy an inhomogeneous evolution equation \cite{tv},
and this fact has been verified at one-loop level
in Ref.\cite{gra}.

Let us consider the case in which
the momentum transfer $t=-(p-p^\prime)^2\ll Q^2$ is also measured.
The current and target fragmentation contributions are in this case
\begin{equation}
\label{currentt}
\sigma_J=\int \f{dx^\prime}{x^\prime} \f{dz^\prime}{z^\prime}
\hsp F_A^i(x^\prime,Q^2)\hsp {\hat \sigma}_{ij} (x/x^\prime,z/x^\prime z^\prime,t x^\prime/z^\prime,Q^2)
\hsp D^j_{A^\prime}(z^\prime,Q^2)
\end{equation}
and
\begin{equation}
\sigma_T=\int \f{dx^\prime}{x^\prime}
\hsp {\cal M}^i_{AA^\prime}(x^\prime,z,t,Q^2)
\hsp {\hat \sigma}_i(x/x^\prime,Q^2)
\end{equation}
The function
${\cal M}^i_{A,A'}(x,z,t,Q^2)$ is the extended fracture function.
In Ref. \cite{gtv} it has been shown that this object can be defined just in
terms of a new cut vertex.
Whereas $M^i_{A,A^\prime}(x,z,Q^2)$ is expected to satisfy an inhomogeneous
evolution equation \cite{tv}, the extended fracture function obeys a simple
DGLAP evolution equation \cite{gtv}.

This facts can be understood by making the following observations.
In the region $\Lambda_{QCD}^2\ll t\ll Q^2$
the hard semi-inclusive cross section
${\hat \sigma}_{ij}$ in eq. (\ref{currentt})
develops large $\log Q^2/t$ corrections
which have to be resummed \cite{graz}.
Instead of absorbing these logs in ${\hat \sigma}_{ij}$ we choose to move them
in the extended fracture function which, by using Jet Calculus, can be defined
in the perturbative region of $t$ as
\begin{align}
\label{defpert}
\!\!\!{\cal M}^j_{A,A'}&(x,z,t,Q^2)= \f{\as (t)}{2\pi t} 
\int F_A^i(w,t)\hsp dw\hsp {\hat P}_i^{kl}(u)\hsp du\hsp
D_{l,A'}(\xi,t)\hsp d\xi \nn\\
&\times E_k^j(r,t,Q^2)\hsp dr\hsp\delta(x-ruw)\hsp \delta(z-w(1-u)\xi)\nn\\
&= \f{\as (t)}{2\pi t} \int \f{dw}{rw}\f{dr}{w-x/r}
F_A^i(w,t)\hsp {\hat P}_i^{kl}(\f{x}{rw})\hsp D_{l,A'}\left(\f{z}{w-x/r},t\right)
E_k^j(r,t,Q^2)\nn\\
&= \f{\as (t)}{2\pi t} \int^{1-z}_x\f{dr}{r}\int^1_{z+r}\f{dw}{w(w-r)}
F_A^i(w,t){\hat P}_i^{kl}\left(\f{r}{w}\right)
\bks D_{l,A'} \left(\f{z}{w-r},t\right) E_k^j(x/r,t,Q^2)
\end{align}

where the integration limits have been obtained implementing momentum
conservation.
We work within the leading logarithmic
approximation.
Here $P_i^j(u)$ and ${\hat P}_i^{jk}(u)$ are regularized and real Altarelli-Parisi vertices \cite{jet}
respectively, 
and the function $E_k^j(x,Q^2_0,Q^2)$ is the evolution kernel
from $Q_0^2$ to $Q^2$,
which obeys the DGLAP evolution equation:
\begin{equation}
Q^2 \f{\d}{\d Q^2} E_i^j(x,Q^2_0,Q^2)= \f{\as (Q^2)}{2\pi}\int_x^1 \f{du}{u}
P_k^j(u) E_i^k(x/u,Q^2_0,Q^2).
\end{equation}
Eq. (\ref{defpert}) is represented in Fig. 1.
\begin{figure}[htb]
\begin{center}
\begin{tabular}{c}
\epsfxsize=8truecm
\epsffile{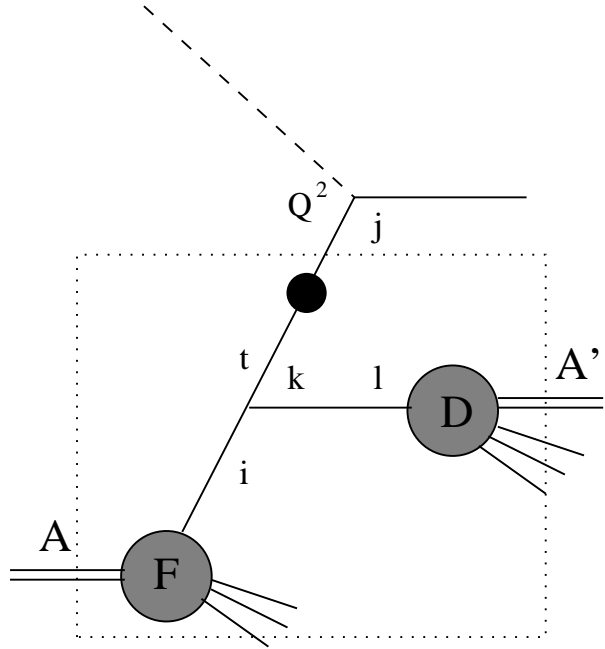}\\
\end{tabular}
\end{center}
\label{1}
\caption{{\em Extended fracture function in the perturbative region of $t$}}
\end{figure}
The evolution kernel (the black blob in Fig. 1) resums the dependence on $Q^2$ and $t$ in terms of logarithms
of the form $\log\f{Q^2}{t}$ and at first order in $\as$ has the following
expression
\begin{equation}
E_i^j(x,t,Q^2)\simeq \delta_i^j\delta(1-x)+
\frac{\alpha_S}{2\pi}P_i^j(x)\; \log\frac{Q^2}{t}. 
\end{equation}

In this way we have absorbed these large logarithms in the
definition of ${\cal M}^i_{A,A'}(x,z,t,Q^2)$ which in the perturbative region of $t$ is resolved in the convolution of the objects contained in the dotted box in Fig 1.

From eq. (\ref{defpert}) it appears that the $Q^2$ dependence of the extended fracture function
is completely given by the evolution
kernel $E_i^k(x,Q^2_0,Q^2)$ and therefore the evolution equation is
\begin{equation}
\label{eqomo}
Q^2 \f{\d}{\d Q^2}{\cal M}^j_{A,A'}(x,z,t,Q^2)= \f{\as (Q^2)}{2\pi}
\int_{\f{x}{1-z}}^1 \f{du}{u} P_i^j(u)
{\cal M}^i_{A,A'}(x/u,z,t,Q^2)
\end{equation}
at least in the perturbative region of $t$, i.e. $t\simge \Lambda_{QCD}^2$.
Therefore
the anomalous dimension associated to
the extended fracture function is the same that controls the evolution
of ordinary structure functions.

Let us suppose now that, as argued in Ref. \cite{gtv}, the evolution equation
(\ref{eqomo})
applies also within the small $t$
region and define the integrated fracture function
as an integral over $t$ up to a $Q^2$-dependent
cut-off of order $Q^2$, say $\eps\hsp Q^2$, with $\eps<1$
\begin{equation}
\label{defin}
M^j_{A,A'}(x,z,Q^2)=\int_0^{\eps\hsp Q^2} dt\hsp {\cal M}^j_{A,A'}(x,z,t,Q^2).
\end{equation}
By taking the logarithmic derivative of eq. (\ref{defin}) we have that
\begin{align}
\label{inhom}
\bksm\bksm
Q^2 &\f{\d}{\d Q^2}M^j_{A,A'}(x,z,Q^2)\!=\!\int_0^{\eps\hsp Q^2}
\bksm\bksm dt \hsp Q^2\f{\d}{\d Q^2}
{\cal M}^j_{A,A'}(x,z,t,Q^2)+\eps\hsp Q^2  
{\cal M}^j_{A,A'}(x,z,\eps\hsp Q^2,Q^2)\nn\\
&= \f{\as (Q^2)}{2\pi}
\int_0^{\eps\hsp Q^2}\bks dt \int^1_{\f{x}{1-z}}\f{du}{u} P_i^j(u)
{\cal M}^i_{A,A'}(x/u,z,t,Q^2)
+\eps\hsp Q^2  {\cal M}^j_{A,A'}(x,z,\eps\hsp Q^2,Q^2)\nn\\
&= \f{\as (Q^2)}{2\pi}\int^1_{\f{x}{1-z}}\f{du}{u} P_i^j(u) M^i_{A,A'}(x/u,z,Q^2)+
\eps\hsp Q^2 {\cal M}^j_{A,A'}(x,z,\eps\hsp Q^2,Q^2).
\end{align}
We see appearing an inhomogeneous term in the evolution equation
of $M^j_{A,A'}(x,z,Q^2)$ which
arises from the $Q^2$ dependence of the upper integration limit.
This inhomogeneous term
depends on the value of ${\cal M}^j_{A,A'}(x,z,t,Q^2)$ in the perturbative
region of $t$ where eq. (\ref{defpert}) holds. By using the boundary
condition
\begin{equation}
E_k^j(x,Q^2,Q^2)=\delta_k^j\delta(1-x)
\end{equation}
on eq. (\ref{defpert}) we get for the inhomogeneous term,
up to $\log \eps$ corrections,
\begin{align}
\label{termineino}
\eps Q^2 &{\cal M}^j_{A,A'}(x,z,\eps\hsp Q^2,Q^2)=\nn\\
&=\f{\as(Q^2)}{2\pi}
\int_x^{1-z}\f{dr}{r}
\bks\int^1_{z+r}\f{dw}{w(w-r)}
F_A^i(w,Q^2) {\hat P}_i^{jl}\left(\f{r}{w}\right) 
D_{l,A'}\bks\left(\f{z}{w-r},Q^2\right)\delta(x/r-1)\nn\\
&= \f{\as (Q^2)}{2\pi}
\int_{z+x}^1 
\f{dw}{w(w-x)} F_A^i(w,Q^2){\hat P}_i^{jl}\left(\f{x}{w}\right)
D_{l,A'}\left(\f{z}{w-x},Q^2\right)\nn\\
&= \f{\as(Q^2)}{2\pi}
\int^{\f{x}{x+z}}_x \f{du}{u} \f{u}{x(1-u)}
F_A^i(x/u,Q^2){\hat P}_i^{jl}(u)
D_{l,A'}\left(\f{zu}{x(1-u)},Q^2\right).
\end{align}

By substituting eq. (\ref{termineino}) in eq. (\ref{inhom})
the evolution
equation for $M^j_{A,A'}(x,z,Q^2)$ appears to be
\begin{align}
\label{eqnonomo}
Q^2 \f{\d}{\d Q^2}&M^j_{A,A'}(x,z,Q^2)=
\f{\as(Q^2)}{2\pi}
\int^1_{\f{x}{1-z}}\f{du}{u} P_i^j(u) M^i_{A,A'}(x/u,z,Q^2)\nn\\
&+ \f{\as (Q^2)}{2\pi}\int^{\f{x}{x+z}}_x \f{du}{x(1-u)}
F_A^i(x/u,Q^2){\hat P}_i^{jl}(u)
D_{l,A'}\left(\f{zu}{x(1-u)},Q^2\right),
\end{align}
that is
exactly the one proposed in Ref.\cite{tv}.

Furthermore the perturbative definition (\ref{defpert})
can be used to derive the evolution with $t$ of the extended fracture function.
In order to make the notation less cumbersome,
it is convenient to define
\begin{equation}
\int_0^1 dz z^m\int_0^{1-z} dx x^n M_{A,A^\prime}^j(x,z,Q^2)=
M^{j,AA^\prime}_{mn}(Q^2).
\end{equation}
One can verify that eq. (\ref{defpert}) becomes
\begin{equation}
\label{defpertm}
{\cal M}^j_{mn}(t,Q^2)=\f{\as (t)}{2\pi t} 
P^{kli}_{nm} F^i_{m+n}(t)
D^k_{m}(t)
E_n^{jl}(t,Q^2)
\end{equation}
where we have defined \cite{jet}
\begin{equation}
P^{kli}_{mn}=\int_0^1 du\hsp u^m(1-u)^n {\hat P}_i^{lk}(u)
\end{equation}
and
\begin{equation}
E_n^{jl}(t,Q^2)=\int_0^1 du\hsp u^n
E_l^j(u,t,Q^2).
\end{equation}
The $t$ evolution equation
contains several terms: the first comes
from the canonical scale dependence of the extended
fracture function, the second from the scale dependence of $\as$, and the third from the scale dependence of the
evolution kernel.
There are then two inhomogeneous terms which follow from the $t$ dependence of structure and
fragmentation functions respectively.
By explicitly deriving eq. (\ref{defpertm}) with respect to $t$ we get
\begin{align}
t\hsp\f{\partial}{\partial t}{\cal M}^j_{mn}(t,Q^2)=
&-\left(\delta^{jp}(1+\beta_0 \as(t))+\f{\as (t)}{2\pi}\hsp A_n^{jp}\right)
\hsp {\cal M}^p_{mn}(t,Q^2)\nn\\
&+\hsp\f{\as^2 (t)}{(2\pi)^2\hsp t} 
\hsp P^{kli}_{nm} \hsp A_{m+n}^{ip}\hsp F^p_{m+n}(t)
\hsp D^k_{m}(t)\hsp
E_n^{jl}(t,Q^2)\nn\\
&+\hsp\f{\as^2 (t)}{(2\pi)^2\hsp t} 
\hsp P^{kli}_{nm} \hsp F^i_{m+n}(t)
\hsp A_m^{kp}D^p_{m}(t)
\hsp E_n^{jl}(t,Q^2)
\end{align}
where
\begin{equation}
A_n^{ji}=\int_0^1 du \hsp u^n P_i^j(u).
\end{equation}

We conclude this section by stressing that
one can define the fracture function in a more general way
as an integral up to an arbitrary $Q^2$-dependent integration limit
\begin{equation}
\label{t2}
M^j_{A,A'}(x,z,Q^2)=\int_0^{t_2(Q^2)} dt~{\cal M}^j_{A,A'}(x,z,t,Q^2)
\end{equation}
In this case the evolution equation reads
\begin{align}
\bksm\bksm
Q^2 &\f{\d}{\d Q^2}M^j_{A,A'}(x,z,Q^2)=
\f{\as(Q^2)}{2\pi}
\int^1_{\f{x}{1-z}}\f{du}{u} P_i^j(u) M^i_{A,A'}(x/u,z,Q^2)\nn\\
&+ \f{\as (t_2(Q^2))}{2\pi}\f{Q^2 t_2^\prime (Q^2)}{t_2(Q^2)}
\int_x^{1-z}\f{dr}{r}\int^{\f{r}{r+z}}_r \f{du}{r(1-u)}
F_A^i\left(r/u,t_2(Q^2)\right){\hat P}_i^{kl}(u)\nn\\
&\times D_{l,A'}\left(\f{zu}{r(1-u)},t_2(Q^2)\right)E_k^j\left(x/r,t_2(Q^2),Q^2\right)
\end{align}
or, by taking double moments,
\begin{align}
Q^2\hsp\f{\partial}{\partial Q^2}M^j_{mn}(Q^2)&=
\f{\as (Q^2)}{2\pi} A_n^{jp} M^p_{mn}(Q^2)\nn\\
+&\f{\as\left(t_2(Q^2)\right)}{2\pi}\f{Q^2\hsp t_2^\prime(Q^2)}{t_2(Q^2)}
P^{kli}_{nm} F^i_{m+n}\left(t_2(Q^2)\right)
D^k_{m}\left(t_2(Q^2)\right)
E_n^{jl}\left(t_2(Q^2),Q^2\right).
\end{align}
In the case in which $t_2(Q^2)=\eps\hsp Q^2$ the
evolution equation (\ref{eqnonomo}) is of course recovered.  

One could even define a more general fracture function
introducing also a lower integration limit $t_1(Q^2)$
\begin{equation}
\label{t1t2}
M^j_{A,A'}(x,z,Q^2)=\int_{t_1(Q^2)}^{t_2(Q^2)} dt~{\cal M}^j_{A,A'}(x,z,t,Q^2).
\end{equation}
and the equation would have an additional inhomogeneous
term taking into account
the $Q^2$ dependence of the lower integration limit
\begin{align}
\bksm
Q^2\hsp\f{\partial}{\partial Q^2}M^j_{mn}(Q^2)&=
\f{\as (Q^2)}{2\pi} A_n^{jp} M^p_{mn}(Q^2)\nn\\
+&\f{\as\left(t_2(Q^2)\right)}{2\pi}\f{Q^2 t_2^\prime(Q^2)}{t_2(Q^2)} 
P^{kli}_{nm} F^i_{m+n}\left(t_2(Q^2)\right)
D^k_{m}\left(t_2(Q^2)\right)
E_n^{jl}\left(t_2(Q^2),Q^2\right)\nn\\
-&\f{\as\left(t_1(Q^2)\right)}{2\pi}\f{Q^2 t_1^\prime(Q^2)}{t_1(Q^2)} 
P^{kli}_{nm} F^i_{m+n}\left(t_1(Q^2)\right)
D^k_{m}\left(t_1(Q^2)\right)
E_n^{jl}\left(t_1(Q^2),Q^2\right).
\end{align}

Phenomenologically these two cases would correspond to the production of hadrons within the target fragmentation region with transverse momentum below
$t_2(Q^2)$ and between $t_1(Q^2)$ and $t_2(Q^2)$ respectively.

\section{Sum rules}
\label{SR}

In Ref.\cite{tv} it was shown that fracture functions obey
the following momentum sum rule
\begin{equation}
\sum_{A^\prime}\int dz\hsp z\hsp M_{A,A^\prime}^j (x,z,Q^2)=(1-x) F_A^j(x,Q^2)
\end{equation}
which accounts for momentum conservation in the $s$-channel.
In this section we want investigate if a momentum sum rule
holds also in the $t$-channel.
Taking moments of eq. (\ref{eqnonomo}) with $m=0$, $n=1$ and summing over $j$
we get \cite{mont}
\begin{equation}
\label{violated}
Q^2 \f{\d}{\d Q^2}\sum_j M^{j,AA^\prime}_{01}(Q^2)=\f{\as(Q^2)}{2\pi}
F_1^{i,A}(Q^2) \int^1_0 du \hsp u \hsp P_i^k(1-u) D_0^{k,A^\prime}(Q^2)
\end{equation}
where we made use of the well-known property of the splitting
function \cite{jet}
\begin{equation}
\sum_j A_1^{ji}=0
\end{equation}
and of the relation
\begin{equation}
\int_0^1 du\hsp (1-u)\sum_k {\hat P}_i^{jk}(u)=\int_0^1 du\hsp (1-u)P_i^j(u).
\end{equation}
From eq. (\ref{violated}) it appears that the derivative of 
$\sum_j M^{j,AA^\prime}_{01}(Q^2)$ receives contribution from the
inhomogeneous
term in the evolution equation. This fact suggests
that the sum rule could be violated because
at values of $t$ of order $Q^2$ the current contribution becomes
important and the two production mechanisms
cannot be disentangled.

As far as the extended fracture function is concerned,
since ${\cal M}^j_{A,A'}(x,z,t,Q^2)$ obeys the DGLAP evolution equation,
it follows that
\begin{equation}
Q^2 \f{\d}{\d Q^2} \sum_j \int _0^{1-z} x\hsp dx
\hsp {\cal M}^j_{A,A'}(x,z,t,Q^2)=0
\end{equation}
and momentum conservation in the $t$ channel is recovered.
In the perturbative region of $t$, by using eq. (\ref{defpert}),
we find
\begin{equation}
\sum_j {\cal M}_{01}^{j,AA^\prime}(t,Q^2)= \f{\as (t)}{2\pi t}
\hsp F^{i,A}_1(t) 
\int^1_0 du\hsp  u P_i^k(1-u) D_0^{k,A^\prime}(t).
\end{equation}

\section{Summary}
\label{SUM}

In this note we showed that a
formulation of
perturbative processes in the target fragmentation region can be given by
using the formalism of Jet Calculus \cite{jet}.

A perturbative evaluation in terms of evolution equation and anomalous
dimensions of extended fracture function is consistent with the
evolution equations proposed in Ref.\cite{tv} and \cite{gtv} for ordinary and
extended fracture functions respectively.
We showed that the inhomogeneous term in eq. (\ref{eqnonomo})
precisely stems from the integration over momentum transfer $t$.

Moreover in this formalism the extension to next-to-leading order
seems quite natural and suggests that eqs. (\ref{eqomo}) and
(\ref{eqnonomo}) keep the same structure in the two-loops
approximation, analogously to what happens for ordinary structure and
fragmentation functions \cite{jet2}.
 
In the region $\Lambda_{QCD}^2\ll t\ll Q^2$
the hard semi-inclusive cross section
develops large $\log Q^2/t$ corrections
which need to be resummed.
Our approach is based on the idea of absorbing such corrections
in the definition of
${\cal M}^i_{A,A'}(x,z,t,Q^2)$,
namely, $\log Q^2/t$ are fully
contained in the evolution kernel $E_i^k(x,t,Q^2)$.
These logs, which can be treated with standard
renormalization group techniques, are new and potentially useful to understand
the dynamics of hadro-production in the target fragmentation region.
These logs could affect significantly
observables like multiplicities and particle distributions and we believe
that a phenomenological study of the role of these corrections would be useful.

As originally proposed in Ref.\cite{tv} and in Ref.\cite{soper},
there now seems to be a widespread consensus
that the $Q^2$ evolution in the
diffractive regime can be
described in terms of the perturbative QCD
formalism \cite{collins,consensus,fs}. 
The data taken by H1
collaboration \cite{h1}, showing an evidence of a consistent
logarithmic scaling violation in diffractive channels have further
stressed that such an interpretation is not far from the experimental
observation. All this
is consistent with
a description in terms of
fracture functions.
Our results agree with such an expectation, although in the limit $z\to 1$
our formulae will show the appearance of large $\log (1-z)$ factors
which should be resummed as well.
For the treatment of such a singular region we
believe an approach can be used close to the one
followed in the inclusive case.

As a concluding remark we want to stress that
the results presented here completely
agree with those of Ref.\cite{gtv}, obtained within
a quite different framework, namely, that
of a generalized cut vertex expansion which, as a generalization of Operator
Product Expansion, appears to be founded on a firmer theoretical standpoint.
Once more the equivalence of parton-like and OPE-like approaches
is verified.
\newpage
\begin{center}
\begin{large}
{\bf Acknowledgments}
\end{large}
\end{center}
\noindent Many of the arguments covered
in this letter originate from
discussions with G. Veneziano.
We also thank G. Veneziano for useful comments,
and S. Catani, D. Graudenz and J. Kodaira for 
conversations. One of us (M.G.) would like to thank the Department of
Physics of the university of Florence for the hospitality at various stages
of this work.

\end{document}